\documentclass[times,authoryear]{elsarticle}

\usepackage{etoolbox}
\makeatletter
\patchcmd{\NAT@citex}
  {\ifNAT@swa\else\NAT@citenum\fi}
  {\ifNAT@swa\else
     \NAT@citexnum
     \ifcase\NAT@ctype\else
       \@ifnum{\NAT@numauthors > 3}{\@nat@et @ }{}
     \fi
   \fi}{}{}
\makeatother

\usepackage{etoolbox}
\makeatletter
\patchcmd{\NAT@citex}
  {\ifNAT@swa\else\NAT@citenum\fi}
  {\ifNAT@swa\else
     \NAT@citenum
     \ifnum\NAT@ctype=0
       \@ifnum{\NAT@numauthors>3}{\NAT@name et al.}{\NAT@name}
     \fi
   \fi}{}{}

\patchcmd{\NAT@bibitem}
  {\if@filesw\immediate\write\@auxout{\string\bibcite{\@citeb}{\the\c@NAT@ctr}}\fi}
  {\if@filesw\immediate\write\@auxout{\string\bibcite{\@citeb}%
    {\the\c@NAT@ctr \ifnum\NAT@numauthors>3 , \textit{et al.}\fi}}\fi}{}{}
\makeatother

\usepackage[]{jasr}
\usepackage{framed,multirow}

\usepackage{amssymb}
\usepackage{latexsym}
\usepackage[switch]{lineno}

\linenumbers                

\let\linenumbers\nolinenumbers\nolinenumbers

\usepackage{url}
\usepackage{xcolor}
\usepackage[colorlinks=true,
            citecolor=blue, 
            linkcolor = blue,
            urlcolor  = blue,
            citecolor = blue,
            anchorcolor = blue]{hyperref}
\definecolor{newcolor}{rgb}{.8,.349,.1}

\usepackage{etoolbox}

\journal{Advances in Space Research}

\begin{document}
\nolinenumbers              

\verso{Lokaveer A \textit{etal}}

\begin{frontmatter}

\title{The SAP-1 Payload: A Technology Demonstration for Space-Based Microbiology Experiments \tnoteref{tnote1}}%
\tnotetext[tnote1]{ SAP-1 stands for SSPACE Astrobiology Payload 1  where SSPACE is Small-spacecraft Systems and PAyload CEnter operated from Indian Institute of Space Science and Technology, Thiruvananthapuram, Kerala, India.}

\author[1]{A \snm{Lokaveer} }

\author[1,2]{Vikram \snm{Khaire}\corref{cor1}}
\ead{vikramk@iittp.ac.in}
\cortext[cor1]{Corresponding author}

\author[1]{Thomas \snm{Anjana}} 

\author[1]{Maliyekkal \snm{Yasir}}
\author[1]{S \snm{Yogahariharan}}
\author[1]{Akash \snm{Dewangan}}

\author[3]{Saurabh Kishor \snm{Mahajan}}
\author[3]{Sakshi Aravind \snm{Tembhurne}}
\author[3]{Gunja Subhash \snm{Gupta}}

\author[1,5]{Devashish \snm{Bhalla}}
\author[1,6]{Anantha Datta \snm {Dhruva}}

\author[4]{Aloke \snm{Kumar}}
\author[4]{Koushik \snm{Viswanathan}}

\author[1]{Anand \snm{Narayanan}}
\author[1]{Priyadarshnam \snm{Hari}}


\affiliation[1]{organization={Indian Institute of Space Science and Technology},
                addressline={Valiamala Road, Valiamala},
                city={Thiruvananthapuram, Kerala},
                postcode={695547},
                country={India}}

\affiliation[2]{organization={Department of Physics, Indian Institute of Technology Tirupati},
                addressline={Venkatagiri road, Yerpedu},
                city={Tirupati, Andhra Pradesh},
                postcode={517619},
                country={India}}
                
\affiliation[3]{organization={Atria University},
                addressline={1st Main Road, AGS Colony, Anandnagar, Hebbal},
                city={Bengaluru, Karnataka},
                postcode={560024},
                country={India}}

\affiliation[4]{organization={Indian Institute of Science},
                addressline={CV Raman Road},
                city={Bengaluru,  Karnataka},
                postcode={560012},
                country={India}}

\affiliation[5]{organization={Vikram Sarabhai Space Centre},
                addressline={Perumathura Road, Veli},
                city={Thiruvananthapuram, Kerala},
                postcode={695022},
                country={India}}

\affiliation[6]{organization={Space Applications Centre},
                addressline={Satellite Road},
                city={Ahmedabad, Gujarat},
                postcode={380015},
                country={India}}
                
\received{}
\finalform{}
\accepted{}
\availableonline{}

\begin{abstract}
The SSPACE Astrobiology Payload (SAP) series, starting with the SAP-1 project 
is designed to conduct in-situ microbiology experiments in low earth orbit. This payload series aims to understand the behaviour of microbial organisms in space, particularly those critical for human health, and the corresponding effects due to microgravity and solar/galactic radiation. SAP-1 focuses on studying \textit{Bacillus clausii} and \textit{Bacillus coagulans}, bacteria beneficial to humans. It aims to provide a space laboratory for astrobiology experiments under microgravity conditions. The hardware developed for these experiments is indigenous and tailored to meet the unique requirements of autonomous microbiology experiments by controlling pressure, temperature, and nutrition flow to bacteria. A rotating platform, which forms the core design, is innovatively utilised to regulate the flow and mixing of nutrients with dormant bacteria. The technology demonstration models developed at SSPACE have yielded promising results,  with ongoing efforts to refine, adapt for space conditions, and prepare for integration with nanosatellites or space modules. The anticipated payload will be compact, approximately 1U in size (10cm x 10cm x 10cm), consume less than 5W power, and offer flexibility for various microbiological studies.
\end{abstract}


\end{frontmatter}

\linenumbers

\section{Introduction}
\label{sec1}

The quest to understand the origins of life on Earth and its possibility elsewhere in the universe has become a frontier theme in humanity's exploration of outer space. Major space organizations, including the Indian Space Research Organisation (ISRO) with its planned \textit{Gaganyaan} human space flight 
program\footnote{\href{https://www.isro.gov.in/Gaganyaan.html}{\tt {https://www.isro.gov.in/Gaganyaan.html}}}, continue to explore in various experimental ways the impact of the hostile conditions that prevail in space (microgravity, ultraviolet flux, and particle radiation field) on biological systems. Understanding the impact of space on microbial life forms has been a dominant theme of research in this direction with several objectives and approaches as summarised in \citet{horneck2010}. In this context, the Small-spacecraft Systems and PAyload CEntre\footnote{\href{https://www.iist.ac.in/sspace}{\tt {https://www.iist.ac.in/sspace}}} (SSPACE)  at the Indian Institute of Space Science and Technology (IIST), Thiruvananthapuram has initiated the SSPACE Astrobiology Payload program (SAP), with SAP-1 as its first mission.  This initiative complements missions like ISRO's Gaganyaan, India's ambitious manned space mission, by providing essential insights into how space travel affects biological entities, a critical aspect for the success and safety of manned missions.

The SAP is a long-term program aimed at developing modular devices for conducting in-situ microbiology experiments in the low earth orbit (LEO). While drawing parallels with NASA's Organism/Organic Exposure to Orbital Stresses (OOREOS) satellite \citep{ooreos}, SAP’s primary objective is to create versatile modules adaptable for small satellites or space platforms aligned with ISRO’s missions. The genesis of this program was inspired by the proposed microgravity experiment \citep{manas}, Microbial Analysis in Space (MANAS) which is focused on studying the growth patterns of \textit{Sporosarcina pasteurii}, a spore-forming bacterium renowned for its ability to induce calcite precipitation.
This foundation paves the way for SAP-1, the initial mission in a series of astrobiology research endeavors by the SSPACE  at IIST. SAP-1 and its successors aim to uncover astrobiological insights and significantly contribute to the success of future human space exploration, leveraging small satellites for a broad spectrum of microbiological studies in LEO.

SAP-1 aims to study the intricate interactions between microorganisms and the harsh conditions of space, particularly microgravity, by employing  optical density (OD) measurement techniques. It focuses on the growth characteristics of probiotic bacteria, specifically \textit{Bacillus clausii} and \textit{Bacillus coagulans}, which have shown significant health benefits but remain relatively unexplored in the context of adverse space conditions of the LEO. The objectives of SAP-1 extend beyond theoretical exploration, aiming to provide tangible insights that can enhance the safety and success of future human space missions. The SAP-1 platform addresses the multifaceted challenges posed to microbiology experiments by outer space conditions, including maintaining optimal pressure and temperature and supplying essential nutrients for bacterial growth. To realize these goals, a novel approach employing a rotating platform with transparent chambers (bio-wells) and a nutrition media reserve has been devised, offering a comprehensive solution to the critical challenges.

The development of a concept demonstration model has yielded promising results, with ongoing efforts to miniaturize and ruggedize the prototypes for integration into nanosatellites or space experimental modules. The envisioned specifications for the final payload include a compact dimension of approximately 10 cm x 10 cm x 10 cm, minimal electrical power consumption (less than 5W), and the versatility to conduct various microbiological experiments across different specimens.

This paper presents the conceptual demonstration model of SAP-1, outlining its motivation, primary payload design, and auxiliary electronics. The structure of the paper is as follows: Section 1 delves into the selection of the biological specimen for SAP-1 and the reasoning behind this choice. Section 2 introduces the concept model of SAP-1, exploring the various mechanical components necessary for executing experiments in LEO. Section 3 addresses the operational electronics essential for performing in-situ measurements and transmitting data to the ground station. Finally, Section 4 provides a summary of SAP-1.


\section{Proposed Microbiology Experiment in SAP-1}

\subsection{Introduction}
Microbiological experiments conducted in LEO serve as pivotal endeavours in advancing our comprehension of how living organisms adapt to the formidable challenges presented by the space environment. LEO is distinguished by a confluence of particle radiations emanating from galactic and solar sources, solar electromagnetic radiations at ultraviolet energies, the absence of atmospheric pressure, substantial temperature oscillations, and microgravity, collectively rendering it an inhospitable milieu for biological systems. The exposure to microgravity, along with ionizing radiation originating from both solar and galactic origins, leads to an adverse space environment detrimental to biological entities. The objectives of space biology experiments encompass the provision of empirical evidence regarding the enduring impacts of space on life systems, as well as the exploration of the broader inquiry into the origins and proliferation of life beyond Earth's confines. The former objective holds direct relevance for both short and long-term human spaceflight programs, as the challenging extraterrestrial conditions expose astronauts to multifaceted physiological and psychological threats \citep{mb5, mb6, mb7}. The influences of reduced gravity and ionizing radiations extend to sub-cellular levels, perturbing gene expression and cellular growth pathways \citep{mb5}.

\subsection{Gut Microbiota}
The human digestive tract harbors a diverse community of over a trillion beneficial microorganisms, each individual exhibiting a unique variance and abundance of these entities, thereby playing integral roles in metabolism, immunity, and nutrition \citep{mb1, mb2}. Research indicates that spaceflight induces compositional alterations in the microbiota of the gastrointestinal tract, skin, nasal passages, and oral cavity, consequently impacting diverse biological functions \citep{mb8}. The gut microbiome significantly contributes to vital physiological processes, including digestion, immune system development, central nervous system maturation, and vitamin production \citep{mb3, mb4}.

Moreover, the health of the gut intricately influences the emotional and cognitive centers of the brain through the communication pathways of the nervous system, collectively known as the gut-brain axis \citep[see e.g.][]{Carabotti15}. Extended periods in space can impose stress on gut homeostasis, the capacity to adapt to changing external conditions, thereby affecting both physiological and psychological well-being. To address such challenges, the oral administration of probiotics emerges as a validated approach for restoring balance to the gut microbiome, enhancing immune function, and sustaining gut health \citep{mb3}.

The potential advantages presented by probiotics underscore the need for extensive research in this relatively unexplored domain. Probiotics, characterized by their diverse benefits and facile administration, emerge as ideal health supplements for astronauts. However, the viability and efficacy of probiotic samples when stored in the space environment for an extended duration remain a relatively unexplored aspect, warranting further investigation.

\subsection{Probiotics in space}
Current research provides insights into the pivotal role of gut microbiota in modulating the gut-brain axis, a bidirectional communication network that establishes a connection between the central nervous system and the enteric nervous system (ENS). This intricate nexus integrates cognitive and emotional brain centers with peripheral intestinal operations, allowing for bidirectional interplay between the microbiota and the gut-brain axis through signaling mechanisms extending from the gut microbiota to the central nervous system and vice versa \citep{mb9}.

The homeostasis of the gut microbiota, crucial for efficient communication between the brain and gut, can be influenced by the space environment, potentially disrupting optimal gut-brain communication. To address this, the administration of probiotics, known for their capacity to harmonize the gut microbiome, emerges as an effective strategy \cite{mb10}.

Various factors in the space environment, including microgravity, seclusion, circadian rhythm changes, and additional stressors, can precipitate psychological disturbances in astronauts, such as anxiety, depression, and related mental health challenges \citep{mb13, mb14}. Emotional well-being is as paramount as physical well-being for astronauts conducting experiments in space. Disruption in microbiota homeostasis can impact mental health by directly influencing the release of neurotransmitters such as serotonin and dopamine, regulating the stress response by affecting the hypothalamus–pituitary–adrenal axis. Microbiota alterations can also influence the levels of BDNF (brain-derived neurotrophic factor) and stimulate inflammatory cytokine release, contributing to the intricate interplay between gut health and mental well-being \citep{mb11}. Empirical research indicates an increased propensity for emotional responses and psychiatric disorders in astronauts during space missions. Clinical investigations, including controlled trials and observational studies on probiotics' application for anxiety and depression management, have demonstrated a reduction in the severity of these neuropsychiatric conditions \citep{mb11, mb12}. Hence, probiotics have the potential to contribute significantly to the mental well-being of space travelers.

In the space environment, astronauts encounter gastrointestinal disturbances, including epithelial barrier disruption and immunological dysregulation, potentially attributed to microgravity's effects on the intestinal epithelium \citep{mb15}. Studies indicate \citep{mb5} an increase in pathogenic potential and antibiotic resistance in specific microorganisms following spaceflight exposure. Certain strains of probiotic bacteria exhibit antimicrobial characteristics, often linked to the secretion of specific peptides or bioactive compounds. These molecules not only confer a competitive advantage within the intricate gut microbiome but also provide host protection against pathogenic bacteria, thereby promoting the persistence of commensal bacteria. Additionally, certain probiotic bacteria strains have been demonstrated to modulate mucin expression, consequently altering the mucus barrier's characteristics and indirectly influencing the gastrointestinal immune response \citep{mb5}.

\subsection{\textit{Bacillus clausii}}
\begin{figure}
\centering
\includegraphics[scale=0.6]{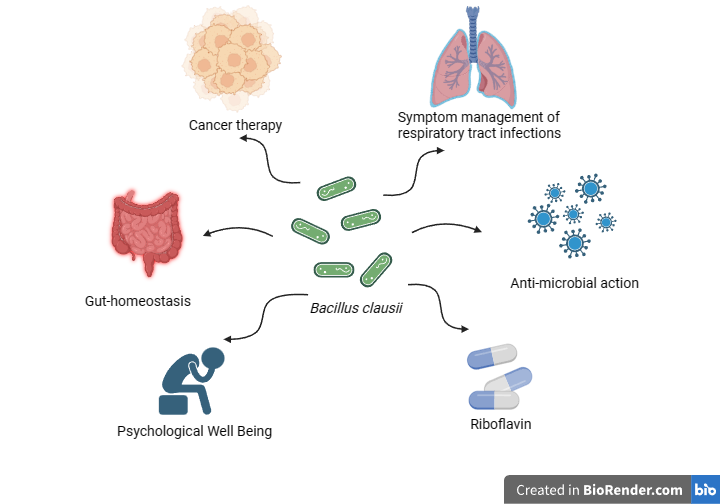}
\caption{Mind-map depicting the various beneficial effects of \textit{Bacillus clausii} (image generated using bioRender). }
\label{fig:benifits}
\end{figure}

The proposed payload is primarily crafted to investigate the effects of the space environment on \textit{Bacillus clausii}, a gram-positive, motile, spore-forming, non-toxic probiotic bacterium. These probiotics encounter formidable challenges as they navigate the harsh conditions of the gastrointestinal tract to reach their target site of action. While the gastrointestinal tract poses a challenging environment for vegetative bacterial cells, spores, renowned for their exceptional resilience, are better suited to withstand these conditions. Notably, these spores undergo germination solely when exposed to favorable conditions in the presence of specific germinants. It is essential to highlight that once germinated, the bacteria lose their resistance to environmental stress. 
There are several beneficial effects of \textit{Bacillus clausii}, including center therapy,  Gut-homeostasis, and symptom management for respiratory tract infections and others as depicted in an illustration given in Figure~\ref{fig:benifits}.
The beneficial effects of probiotic bacteria are contingent upon the germination of spores at the intended location in the gut. Therefore, it is imperative to scrutinize the growth behavior of these bacteria in a space environment to optimize their efficacy in space applications. Although \textit{Bacillus clausii} is relatively less studied, it exhibits SAP-1 in conferring numerous health benefits.

Upper respiratory tract infections stand among the most prevalent infections in pediatric populations. Research suggests positive outcomes associated with probiotics in treating these infections, particularly the efficacy of \textit{Bacillus clausii} UBBC-07 in managing upper respiratory tract infection symptoms \cite{mb16, mb17}. A key rationale for administering \textit{Bacillus clausii} as a probiotic during antibiotic regimens lies in its intrinsic resilience to a broad spectrum of antibiotics. Consequently, it retains the ability to populate the gastrointestinal tract even in the presence of reduced commensal bacteria resulting from antibiotic exposure, thus aiding in mitigating the risk of gut dysbiosis. \textit{Bacillus clausii} secretes compounds with antimicrobial activity against diverse pathogens, playing a pivotal role in maintaining a harmonious gastrointestinal microbial equilibrium \cite{mb18, mb19}.

The emerging consensus on the significant role of gut microbiota in modulating responses to cancer treatments has prompted attention. A robust and diverse gut microbial composition is believed to enhance outcomes of immunotherapy and chemotherapy \cite{mb21, mb22}. Many gut bacteria, including \textit{Bacillus clausii}, produce metabolites with observed anti-tumorigenic properties. \textit{Bacillus clausii} is part of a distinct intra-tumoral microbial signature associated with prolonged survival in patients with surgically resected pancreatic adenocarcinoma, along with \textit{Pseudoxanthomonas}, \textit{Streptomyces}, and \textit{Saccharopolyspora} \cite{mb20}. The spores of \textit{Bacillus clausii} UBBC-07 have also demonstrated benefits in ameliorating oral mucositis symptoms in patients with head and neck cancer undergoing radiation treatment \cite{mb23}.

Recognition of the gut-brain axis and the potential therapeutic role of probiotics is growing for enhancing interventions in stress and various neurological conditions. \textit{Bacillus clausii} has demonstrated the ability to alleviate stress-related behavior in mice through augmentation of monoamine levels, along with an increase in the mRNA expression of dopamine receptors (D1 and D2) and synaptophysin \cite{mb24}. Despite the essential role of riboflavin in cellular functioning and growth, humans do not endogenously synthesize it. Specific strains of \textit{Bacillus clausii} can bio-synthesize and secrete riboflavin, potentially contributing to enhanced cellular functionality and growth in humans \cite{mb17}. A comprehensive analysis of the entire genome sequence of \textit{Bacillus clausii} UBBC07 has revealed the presence of antibiotic-resistance genes within its chromosomal DNA, which are intrinsic and not transferable. Additionally, no toxin genes were detected, collectively indicating that the consumption of \textit{Bacillus clausii} UBBC07 is considered safe for human consumption \cite{mb25}.

\subsection{SAP-1 payload with Bacillus clausii}
Probiotics, emerging as a promising intervention for diverse challenges encountered during space flight, represent an exemplary subject for research and exploration. Given their biological nature, comprehensive investigations into the behavior and viability of probiotics in the space environment are imperative for tailoring their administration in future human space missions. The selected probiotic bacterium, \textit{Bacillus clausii}, is ingested in the form of spores. Endospores, characterized by their high resistance and thick-walled structures formed within bacterial cells as a survival strategy in adverse conditions, germinate into live bacteria upon exposure to favorable conditions. The planned experiments for the proposed payload aim to study the revival of \textit{Bacillus clausii} spores into live bacteria and their subsequent proliferation.

\begin{figure}[h]
\centering
\includegraphics[scale=0.6]{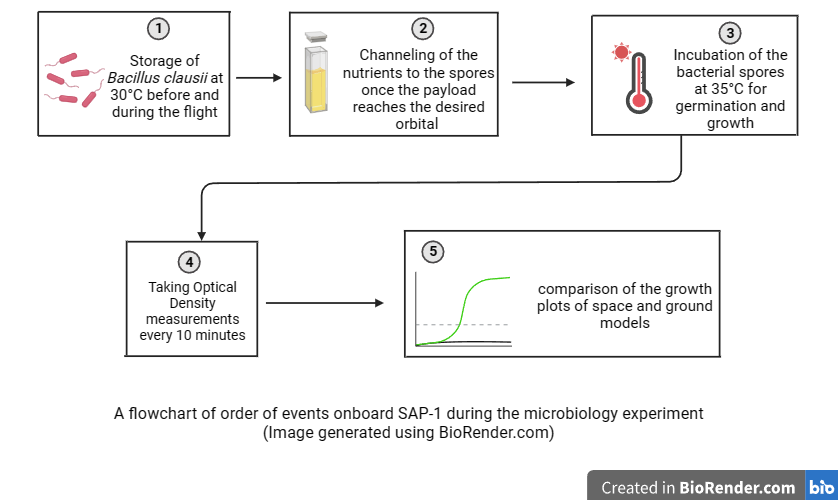}
\caption{A flowchart of the order of events to be conducted onboard SAP-1 during the microbiology experiment (image generated using bioRender). }
\label{fig:eventflow}
\end{figure}

Utilizing the SAP-1 bio hardware developed by SSPACE at IIST, dry bacterial spores can be transported to space in small bio-wells, accompanied by the provision of supplying nutrient broth at the designated activation time (Figure \ref{fig:eventflow}).  The current version of the SAP-1 has four bio-wells, however, this number can be easily increased in the final space-grade version.  Growth rate tracking, facilitated by OD measurements, is a valuable tool for comparing space-based outcomes with ground-based control experiments and deciphering microbial growth in the space environment. The storage of \textit{Bacillus clausii} spores at 30°C, followed by incubation at 35°C for germination and culturing, is the proposed methodology. Once the culture has saturated and reached the senescence phase, the experiment will be terminated by heating the payload to over 120°C, thus preventing any kind of biological contamination in space. Once the culture has saturated and reached the senescence phase, the experiment will be concluded by elevating the payload's temperature to over 120 degrees Celsius. Similar measures are employed in the event of an intermediate termination of the experiment. This is implemented to prevent any potential biological contamination in the space environment effectively.

To ensure the unhindered growth of bacterial cultures in the bio-well, it is essential to ascertain the optimal ratio of air to the culture volume. This is achieved through a series of ground-based tabletop experiments, which are elaborated upon in the following subsections.

\subsection{Experiments to determine optimal aeration}

Bio-wells on SAP-1 are air-tight chambers and hence the apt ratio of the quantity of atmospheric air to the quantity of bacterial culture must be known for optimal growth of bacteria. An experiment was carried out using \textit{Bacillus clausii} and \textit{Bacillus coagulans} to determine a favourable ratio by varying aeration levels and quantity of strain cultures and utilizing a closed 24-well plate and falcon tubes for inoculating different culture ratios, monitoring and recording OD measurements over a defined period.

In the experimental methodology, cultures were cultivated in standard Nutrient Broth. The preparation of cultures involved utilizing a 5 mg/ml spore suspension as an inoculum, with inoculation performed using a 10\% suspension of the final volume. The varied volumes of culture across samples were set at 25\% (600 ul), 50\% (1200 ul),  and 95\% (2300 ul) of the total well volume, representing 75\%, 50\%, and 5\% of air volume, respectively. Triplicate samples were grown. Cultures were incubated in a multi-well plate reader for 16 hours at 37℃ and 180 rpm. OD measurements at 600 nm were taken every 30 minutes.

To address path length discrepancies resulting from different culture volumes, a calibration step was introduced. This involved measuring the OD of equivalent volumes from a single saturated culture. Since OD was not directly proportional to sample volume (path length), actual OD readings were normalized using the corresponding calibration factor.

\begin{figure}[h]
\centering
\includegraphics[scale=0.6]{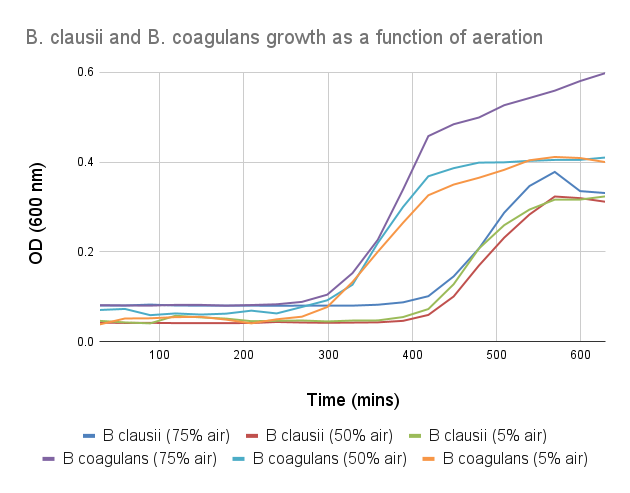}
\caption{Averaged growth curve of \textit{Bacillus coagulans} and \textit{Bacillus clausii} obtained for 5\%, 50\% and 75\% volume of air in bio-wells during experimentation.}
\label{fig:growth}
\end{figure}

Figure \ref{fig:growth} shows the OD measurements for \textit{Bacillus coagulans} and \textit{Bacillus clausii} for 5\%, 50\% and 75\% volume of air in bio-wells. Key observations included:

\begin{enumerate}
    \item \textit{Bacillus coagulans} demonstrated a shorter lag phase, likely due to a higher number of active spores in the initial samples (refer to Figure \ref{fig:growth}).
    \item Cultures with a higher aeration volume (75\%) showed greater final growth density (refer to Figure \ref{fig:growth}).
    \item The growth rates, however, seemed unaffected by the varying aeration volumes (refer to Figure \ref{fig:growth}).
\end{enumerate}

Our experiments indicate that an aeration volume of 75\% is optimal for obtaining OD measurements over a relatively extended period.
While this 75\% optimal aeration volume is based on ground tests, microgravity may slightly alter the required volume due to changes in fluid dynamics - specifically the absence of gravitational convection. However, the chosen volume is sufficient to meet the needs of the initial bacteria and nutrients.

\subsection{Summary of microbiology part of the SAP-1}
In the realm of space exploration, microbiological experiments conducted in LEO are crucial for understanding how living organisms adapt to the challenges posed by the space environment. LEO's unique conditions, including microgravity, radiation, and temperature fluctuations, make it an inhospitable setting for biological systems. These experiments aim to provide empirical evidence on the enduring impacts of space on life systems, directly relevant to human spaceflight programs.

One area of focus is the impact of spaceflight on gut microbiota, a diverse community of microorganisms crucial for metabolism, immunity, and nutrition. Extended space missions can induce compositional changes in the microbiota, affecting physiological processes and potentially disrupting the gut-brain axis.  Probiotics, particularly \textit{Bacillus clausii}, emerge as promising supplements to address these challenges.

Research indicates that probiotics play a vital role in harmonizing the gut microbiome, essential for efficient communication between the brain and gut. In the space environment, factors like microgravity and stressors can lead to psychological disturbances in astronauts. Probiotics have shown SAP-1 in managing anxiety and depression, contributing to the mental well-being of space travelers. Moreover, gastrointestinal disturbances, including epithelial barrier disruption and immunological dysregulation, are observed in astronauts during spaceflight. Probiotic bacteria, with antimicrobial characteristics, can offer protection against pathogenic bacteria, promoting the persistence of beneficial bacteria in the gut.

The proposed payload focuses on investigating the effects of the space environment on \textit{Bacillus clausii}, a resilient probiotic bacterium. \textit{Bacillus clausii} has demonstrated efficacy in managing upper respiratory tract infections, maintaining gut microbial equilibrium during antibiotic regimens, and potentially influencing responses to cancer treatments. Its ability to alleviate stress-related behavior and bio-synthesize riboflavin further enhances its significance.

The SAP-1 Payload aims to explore the behavior and viability of \textit{Bacillus clausii} spores in the space environment. These spores, known for their resistance, germinate into live bacteria under favorable conditions, and the experiments seek to elucidate this process. The comprehensive analysis of \textit{Bacillus clausii}'s genome ensures its safety for human consumption.

Ground-based experiments have been conducted to observe the growth of \textit{Bacillus clausii}. Through various trials, it has been determined that maintaining a 75\% aeration volume in the culture environment results in a more significant final growth. This setup provides a reliable range for OD (OD) measurements over an extended duration.

In summary, the study of microbiological aspects, particularly the role of probiotics like \textit{Bacillus clausii}, is essential for addressing the challenges posed by spaceflight conditions. Understanding the behavior of these probiotics in space can pave the way for tailored interventions to support the health and well-being of astronauts in future human space missions.

\section{Mechanical instrumentation of SAP-1 }

\subsection{Overview}
\begin{figure}[h]
\centering
\includegraphics[scale=0.5]{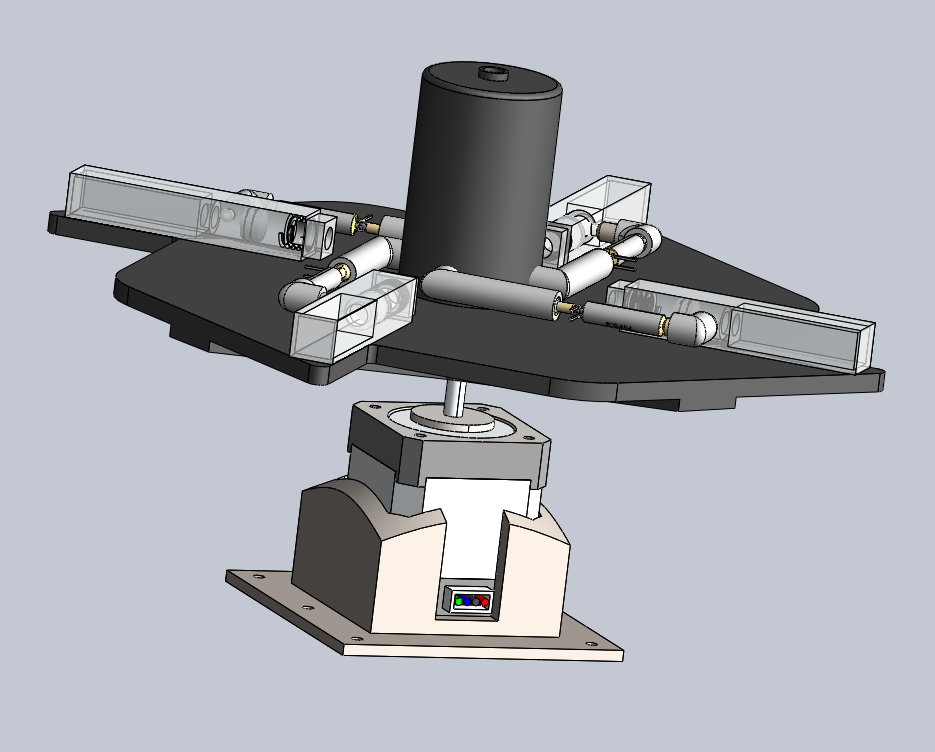}
\caption{Isometric side view of the rotating platform of SAP-1 mounted on the stepper motor encased in the motor mount (CAD model).}
\label{fig:rotside}
\end{figure}

The mechanical instrumentation encompasses the comprehensive configuration of the SAP-1 payload, addressing structural, thermal, and logistical considerations. SAP-1 autonomously executes essential experimental procedures in a distinctive yet straightforward manner. The system features a rotating platform, characterized by axes of symmetry passing through the center and connected to a motor (refer Figure \ref{fig:rotside}). Positioned at the distant ends of the rotating platform are bio-wells containing bacteria in their sporulated form. In the center of the platform, a tank holds the required nutrient broth for spore germination and growth. The tank is maintained at higher pressure, with a piston segregating pressurized air from the nutrient liquid. 
The bio-wells are connected to the tank through solenoid valves and plumbing.
  The nutrient tank is pressurized slightly above atmospheric pressure, while the bio-wells are kept at atmospheric pressure, creating a pressure difference between the two.
To ensure reliable separation of the air and nutrient broth, a piston-based system has been selected due to its ability to provide the necessary displacement force and its proven leak-proof performance under extreme conditions. This design will undergo rigorous testing for leaks to ensure reliability in space-grade applications.
The entire setup is housed within a hermetically sealed chamber, employing appropriate heating and temperature control techniques to maintain Earth-like pressure and temperature conditions.

Upon command, the platform initiates rotation, and valves open for a short, predetermined interval, enabling a specific amount of nutrient broth to flow into the bio-wells. Subsequently, the platform executes an agitation cycle (jittery motion) to ensure a uniform mixing of bacteria with the nutrient medium. During the growth phase, continuous stirring of the bacterial culture is crucial, necessitating similar motor actions for periodic mixing of the cultured media. The lack of gravity in LEO poses challenges in obtaining pure OD (OD) readings of a liquid-air mixture floating in space. To address this, the platform is given a constant angular velocity during the OD Measurement instants, inducing centrifugal forces to force the culture at the farthest end of the bio-well. The hermetically sealed chamber, maintained at 1 bar pressure, is filled with  nitrogen gas. The rotation of the platform also aids in uniformly distributing heat throughout the chamber.

\subsection{The rotating platform}
The rotating mechanism of the platform serves as a comprehensive solution to numerous challenges encountered in automating laboratory-like experimental setups, particularly in a space environment. Several crucial considerations are paramount when replicating Earth-based microbiological experiments in space, as outlined below:

\begin{enumerate}

    \item Ensuring uniform mixing of the bacterial sample with the nutrition broth at the experiment's commencement by the jittery motion of the platform. Homogeneous bacterial culture is imperative for consistent and healthy bacterial growth.

    \item Facilitating intermediate agitation of the bacterial culture to prevent bacterial lumping during growth. This action prevents bacterial colonies from adhering to the surface of the bio-wells, which could otherwise impede OD measurements. The "sloshing" action induced by the platform's jittery rotation ensures a nearly uniform mixing of the culture.    
    \item Implementing density-based separation of the air-culture mixture during OD measurements due to the absence of gravity. This prevents air pockets from interfering with OD measurements and yielding inaccurate readings.

    \item Ensuring the uniform distribution of heat within the hermetically sealed chamber. Given the absence of gravity and the limitations of heat transfer by convection, the rotating motion of the platform introduces forced mechanical convection, promoting uniform heat transfer.

    \item Incorporating a redundant passive pump for the nutrient liquid. In the event of pressure leaks or self-actuation failures in the tank and bio-wells, the rotating action itself generates sufficient centrifugal force to pump liquid from the tank to the bio-wells, albeit at a slower rate.

\end{enumerate}

One of the challenges with using a rotating platform is that it imparts centrifugal acceleration, creating artificial gravity. To address this, the rotation speed must be carefully designed to ensure the average centrifugal acceleration remains within the microgravity range. In our final planned setup, a platform with a diameter of approximately 10 cm
rotating at 5 RPM generates a centrifugal acceleration of the order of 0.01 m$s^{-2}$. However, since the platform rotates intermittently only for the OD measurements lasting about 5 seconds once per hour, the time-averaged centrifugal acceleration is reduced to the order of $10^{-6}g$ where $g$ is the acceleration due to gravity. This setup ensures that the induced centrifugal acceleration remains consistent with microgravity conditions.

\subsection{A brief insight}
\begin{figure}[h]
\centering
\includegraphics[scale=0.6]{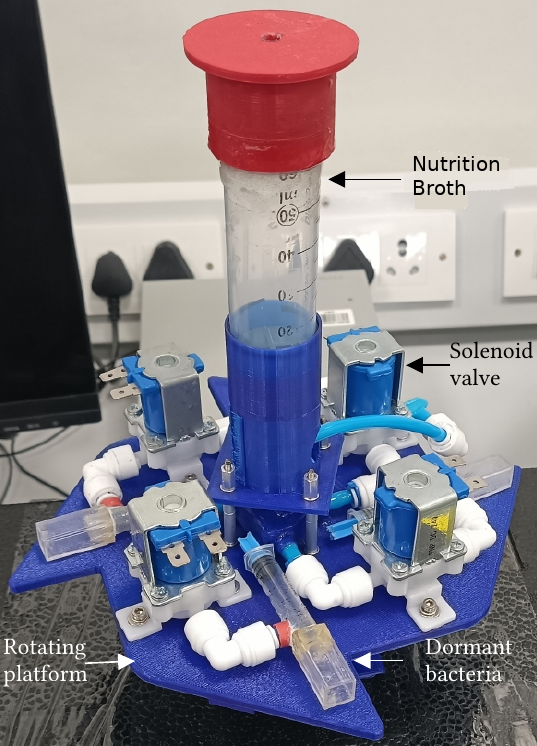}
\caption{Isometric view of the technology demonstration model of the rotating platform with nutrition tank, solenoid valves, bio-wells and plumbings (3D printed tabletop model).}
\label{fig:sap2b}
\end{figure}

The rotating platform within SAP-1 plays a pivotal role by housing essential mechanical components vital for the autonomous execution of experimental procedures. These components include the nutrition broth storage tank, solenoid valves, bio-wells, corresponding plumbing, OD sensors, bio-well heaters, temperature sensors, and the slip ring (refer Figure \ref{fig:sap2b}).

\textbf{The nutrition tank} serves as the reservoir for the nutrition broth required for experimentation within the payload. The tank is presently configured to be pressurised, employing a piston to segregate compressed air from the nutrition broth, ensuring that only the nutrition broth is pumped into the bio-wells.

\textbf{The plumbings and connectors} play a crucial role in interlinking the Tank-Valve-Biowell system, ensuring the containment of the nutrition broth flow within the closed system. Consequently, these plumbings and connectors must be airtight. The tank is configured with a single outlet, later divided into three outlets using a 4-way connector.

\textbf{Solenoid valves} are utilised in SAP-1 due to their capability to open and close upon command. These valves serve as the control interface between the tank and the bio-wells.

\textbf{Bio-wells}, situated on the farther end of the rotating platform, serve as chambers designed to house hibernating/sporulated bacteria, serving as the sites for the actual experimentation within the payload. Consequently, they are constructed with transparency to facilitate OD measurements.

\begin{figure}[h]
\centering
\includegraphics[scale=0.6]{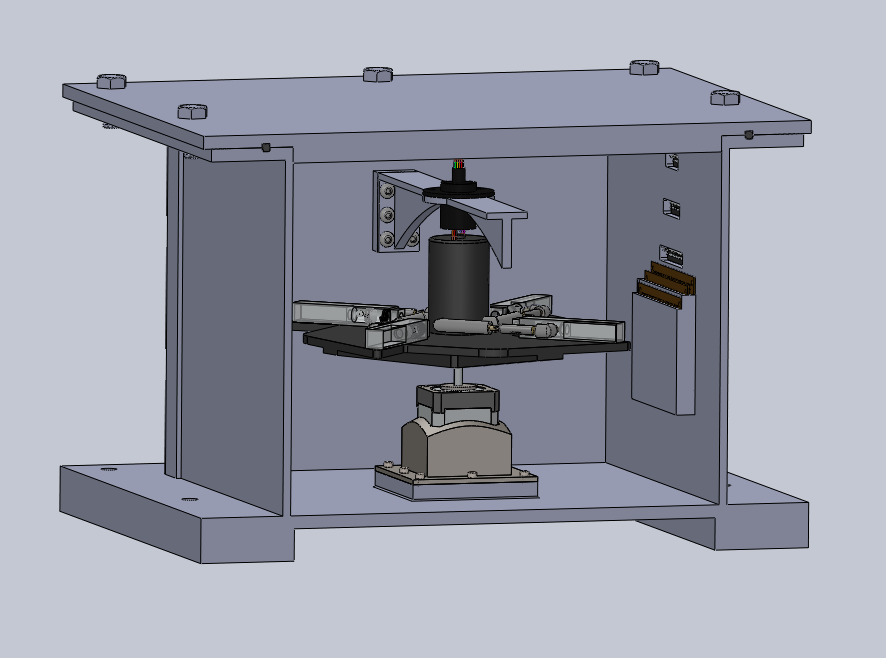}
\caption{Vertical cross-section of the SAP-1 payload encapsulated in a hermetically sealed chamber with a slip ring to maintain electrical contact between the stationary chamber and rotating platform, slip ring supports to support the slip ring, electronic printed circuit boards stacked on the right side and corresponding interfacing ports above the electronics stack (CAD model).}
\label{fig:sap3b}
\end{figure}

The complete rotating platform is encapsulated within a hermetically sealed chamber, which functions as the foundation for the rotating platform and serves as the core support structure for the entire SAP-1 payload (refer Figure \ref{fig:sap3b}). The final dimensions of the entire setup are targeted to be 1U (10cm X 10cm X 10cm) with its weight being under 1500 grams.

\subsubsection{Tests, simulations and analyses}
Structural analysis and thermal analysis of the technology demonstration model of the SAP-1 payload were conducted using simulation software named SOLIDWORKS (by  Dassault Systèmes) and Ansys Thermal Desktop respectively, with satisfactory results. 


\subsection{Summary - Mechanical instrumentation }
The mechanical instrumentation of the SAP-1 payload involves a sophisticated design addressing structural, thermal, and logistical aspects. It employs a rotating platform mechanism with bio-wells containing bacteria, a nutrient tank under pressure, and a hermetically sealed chamber with temperature control for Earth-like conditions in space.

The platform's rotation initiates experimental procedures, ensuring uniform mixing of bacteria with nutrient broth and preventing bacterial lumping. In the absence of gravity, the platform's motion induces centrifugal forces, aiding in obtaining accurate OD measurements. 
 The rotating platform, with its large surface area, helps to mix the air inside the hermetically sealed chamber. This induces forced mechanical convection, allowing heat to be distributed more uniformly by promoting air movement and mixing within the chamber.

The rotating platform in SAP-1 addresses challenges in space experiments by:
\begin{enumerate}
    \item Ensuring uniform bacterial mixing for consistent growth.
    \item Preventing bacterial lumping through intermittent agitation.
    \item Implementing density-based separation for accurate OD measurements.
    \item Promoting uniform heat transfer in the sealed chamber.
    \item Incorporating a redundant passive pump for nutrient circulation.
\end{enumerate}

Essential components within the rotating platform include the nutrient tank, solenoid valves, bio-wells, plumbing, OD sensors, bio-well heaters, temperature sensors, and a slip ring. The tank, pressurized with a piston, provides the nutrition broth, while solenoid valves control the flow to bio-wells. Bio-wells, transparent for OD measurements, house bacteria for experimentation.

The hermetically sealed chamber, enclosing the rotating platform, serves as the foundation and core support for the SAP-1 payload. Structural and thermal analyses of the technology demonstration model have yielded satisfactory results, confirming the payload's robust design.


\section{Electronics instrumentation }

\subsection{Overview}
\begin{figure}[h]
\centering
\includegraphics[scale=0.5]{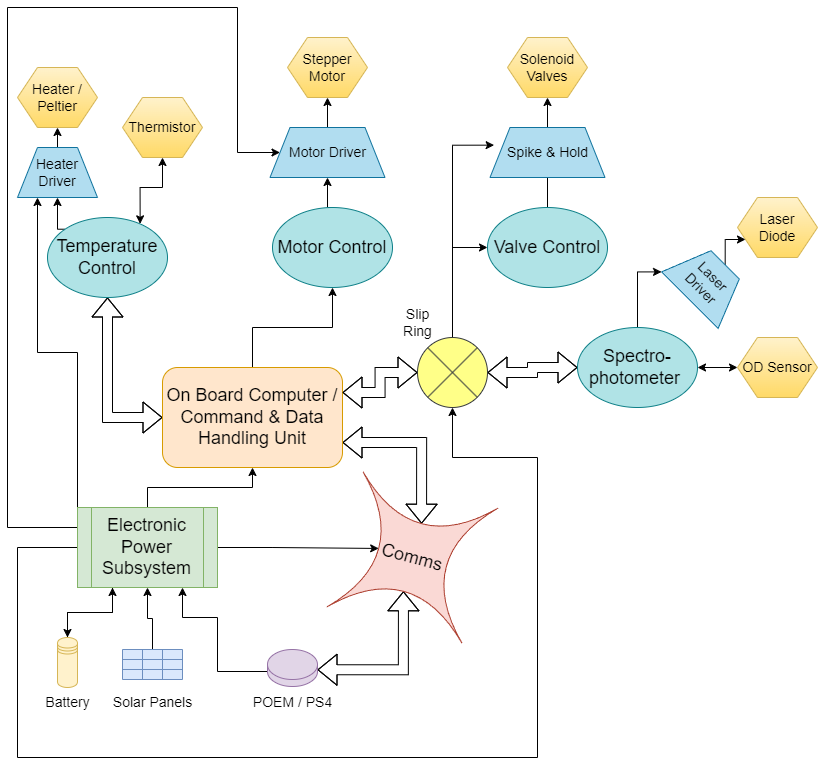}
\caption{Block diagram of the complete electronics and instrumentation system of SAP-1 (created using draw.io).}
\label{fig:sap5a}
\end{figure}

The electronics instrumentation serves as the backbone of the entire operational framework of SAP-1 (refer Figure \ref{fig:sap5a}). It plays a crucial role in commanding and controlling the payload's subsystems, as well as facilitating the relay of information to and from ground control. The primary functionalities of the electronics  can be enumerated as follows:
\begin{enumerate}
    \item Controlling and commanding SAP-1
    \item Accurately executing the autonomous experimentation sequence in SAP-1
    \item Handling the data generated by SAP-1
    \item Communicating the data generated to and accepting manual commands from the ground station
    \item Acquiring, storing, and redistributing electrical power as per requirements and availability
\end{enumerate}

The electronics of SAP-1 can be classified into the following parts:
\begin{enumerate}
    \item Control subsystem
    \item Power subsystem
    \item Communication subsystem
\end{enumerate}

\subsection{Control subsystem}

\textbf{Motor control system:} The rotating platform, serving as the central component of SAP-1's experimental setup, necessitates a stable and robust control system. This control system is essential to execute a diverse range of specifically modelled operation modes and correspondingly regulate the motor.

\textbf{Valve control system:} The valves employed in SAP-1 are solenoid valves, and due to the inductive nature of solenoids, they introduce some transience to step inputs. Consequently, to instantaneously activate a solenoid valve, the magnetic flux of the solenoid's inductor must be rapidly 'magnetically charged' at the moment of the step input and sustained at a constant level after reaching the threshold. Achieving this requires a specialized circuit known as a 'Spike and Hold' circuit. 
 This is an electronic circuit that delivers a high-voltage pulse (spike) to a solenoid valve for a brief duration to activate it, followed by a lower voltage (hold) to maintain its position. This method optimizes power consumption and extends the lifespan of the solenoid valve. The spike voltage overcomes the initial resistance of the valve, while the hold voltage ensures continuous operation without excessive power dissipation. In the current prototype, the spike voltage is 5V, and the hold voltage is 3.3V.

\textbf{Spectrophotometer} is an instrument designed to measure the amount of light absorbed by a sample. Its operation involves passing a light beam through the sample to assess the light intensity. Comprising two main components—a spectrometer and a photometer—the spectrometer generates light of any wavelength, while the photometer gauges light intensity.

In the spectrophotometer setup, the liquid or sample is positioned between the spectrometer and the photometer. The photometer assesses the amount of light traversing the sample, delivering a voltage signal to the display. A change in light absorption corresponds to a change in the voltage signal, aligning with Beer Lambert's Law.

In microbiological applications, OD measurements are typically conducted at a wavelength of 600 nm. This specific wavelength is chosen because most substances involved in microbiological experiments, including microorganisms, exhibit minimal response or interference with light in the vicinity of 600 nm. Consequently, OD at 600 nm is a widely utilised metric in microbiological experiments \cite{mb26}.

Professional spectrophotometers, though effective, are often bulky and impractical for direct integration into application-specific setups like SAP-1. To address this, a miniaturized version of a spectrophotometer has been developed. This miniaturized version incorporates a commercially off-the-shelf high dynamic range digital light sensor, characterised by a vast dynamic range, programmable gain, and an optimised peak.

\textbf{The Temperature control system} is implemented to maintain optimal temperatures required for bacterial growth (ranging from 33 to 37 degrees Celsius) both within the payload and within the bio-wells. Thermistors function as temperature sensors, while resistive heaters serve as the heat sources.

\textbf{The On-board computer (OBC) / Command and data handling system} aboard SAP-1 is designed to execute all the functionalities outlined throughout and ensure seamless coordination among multiple subsystems. Additionally, the OBC is entrusted with the responsibility of safe storage of data generated on-board SAP-1. For this mission, flight-tested (TRL-9) OBCs based on Microsemi SmartFusion-2 processors are employed.

\subsection{Power subsystem}
The power subsystem in SAP-1 serves functions akin to those found in the majority of satellites or payloads but is tailored for shorter-duration operations (around 30 Days). It is tasked with providing power to all electronic subsystems of SAP-1, which is designed to have a peak power consumption of less than 5 watts and a nominal power consumption of 5 watts. As depicted in the block diagram at the beginning of the chapter, the power subsystem receives input power from two sources: one from the power supply of the orbital experimental module and the other from solar panels. In instances of intermittent power supply from the orbital experimental module or when it fails to meet the peak power requirements of SAP-1, a battery will be incorporated to ensure a balanced and consistent power supply. Thus the power subsystem incorporates a battery management system as well.

\subsection{Communication subsystem}
The primary function of the communication subsystem is to reliably transmit the data generated onboard SAP-1 to the ground station upon command. Additionally, it is designed to accept commands from the ground station, enabling the triggering of critical functions such as the initiation and termination of the experiment. The communication systems onboard the orbital experimental module will be employed in conjunction with an in-house developed Ultra High Frequency (UHF) communication module (433MHz) or a Commercially Off-The-Shelf (COTS) S-Band communication module (2.49 GHz).

\subsection{Summary - Electronics instrumentation subsystem}
The electronics serves as the operational backbone of SAP-1, commanding subsystems, and facilitating information exchange with ground control. It encompasses the Control, Power, and Communication subsystems.

Control subsystem:
\begin{enumerate}
 \item Motor control system: Regulates the rotating platform for diverse operation modes.
 \item Valve control system: Utilizes `Spike and Hold' circuits for solenoid valve activation.
 \item Spectrophotometer: Miniaturized version with a Digital Light sensor for OD measurements.
 \item Temperature control system: Maintains optimal temperatures using thermistors and resistive heaters.
 \item On-Board computer: Based on Microsemi SmartFusion-2 processors, ensures seamless coordination and data storage.
\end{enumerate}

Power subsystem:
Tailored for short-duration operations, it provides power from orbital experimental modules and solar panels. A battery ensures consistent supply during power fluctuations.

Communication subsystem:
It transmits data to the ground station and accepts commands. Utilizes UHF or COTS S-Band communication modules for reliable communication.

The electronics is vital for SAP-1's autonomy, experimentation, data handling, and communication. The systems have been tailored for space conditions, ensuring reliable performance.

\section{Conclusions}
The SAP-1 initiative is dedicated to investigating the effects of microgravity and space radiation on living organisms, with a specific focus on the critical role of microbial organisms in life support. SAP-1 concentrates on the probiotic bacteria \textit{Bacillus clausii}, aiming to explore its growth and survival in the space environment at LEO. The significance of this study is rooted in its potential contributions to astronaut health, considering the gut microbiota's essential involvement in digestion, immune function, and mental health.

The importance of probiotics in space becomes apparent as we investigate their ability to alleviate the impacts of space-induced stress, gastrointestinal issues, and psychological challenges encountered by astronauts. The use of \textit{Bacillus clausii}, a robust probiotic strain, emerges as a promising approach to preserving gut health, improving immune function, and potentially supporting mental well-being during prolonged space missions. The payload design within the SAP-1 initiative, incorporating a rotating platform, tackles crucial challenges in automating microbiological experiments in microgravity. This design ensures uniform mixing, prevents bacterial clumping, and facilitates density-based separation for precise OD measurements.

The rotating platform within the mechanical instrumentation plays a crucial role in preserving the integrity of the experiments. It ensures even distribution of nutrient broth and uniform heat transfer within the hermetically sealed chamber. This innovative solution offers a comprehensive approach to overcoming challenges specific to microbiological experiments in space.

The electronics instrumentation functions as the operational core, by controlling the autonomous execution of experimental procedures, managing motor and valve systems, and handling data acquisition and communication with ground control. The integrated control, power, and communication subsystems guarantee the smooth operation of SAP-1, showcasing a careful integration of technology to align with the goals of this pioneering astrobiological research.

 Although the SAP-1 payload can be flown either on an independent nanosatellite or onboard a space platform, the latter option is more advantageous. This approach eliminates the need for controlling the satellite's maneuvers, and the required power can be drawn directly from the space platform. Specifically, we plan to use SAP-1 on the Polar Satellite Launch Vehicle (PSLV) Orbital Experimental Module (POEM)\footnote{More info on POEM: \href{https://www.isro.gov.in/POEM-3_Mission_achieves_Payload_objectives.html}{Webpage link}}.

In conclusion, the SAP-1 initiative, with its interdisciplinary approach by spanning the fields of microbiology, mechanics, and electronics, not only enhances our comprehension of probiotic behavior in space but also establishes a foundation for forthcoming missions involving microbial experiments. As we explore the uncharted research into microgravity, the information gathered from SAP-1 is expected to have a meaningful impact on the well-being and success of astronauts in upcoming space exploration missions.

\section*{Acknowledgments}

We would like to express our sincere gratitude to SSPACE, IIST, Indian Space Research Organisation (ISRO), Department Of Space (DOS), Atria University, IISc, and National Center for Biological Sciences (NCBS), Bengaluru for their invaluable support. Their commitment to advancing knowledge and fostering research excellence has been instrumental in the realization of our goals. Each of these organizations has played a crucial role in providing financial, logistical, and infrastructural support, enabling us to conduct rigorous experiments, analyze data, and disseminate our findings. Their collaborative spirit and dedication to scientific advancement have significantly contributed to the quality and depth of this research. We immensely thank and give due credit to the rest of our team members from IIST, namely Annie Gabriel, Raahil Rana, Hritik Singh Parmar, Rakshita Alandikar, Murali Krishna, Prabhash Singh, Anurag Meshram and Vidish S. We also gratefully thank Deepa Agashe, Pratibha Sanjenbam and Asha from NCBS, Raghav V, Akash Mahobe, Soumyadip Bhunia, Aditya Shukla, Pranav Koppa, Darshil Sojitra and Manoj Agrawal from IIST for their support.

\section*{Declaration of generative AI and AI-assisted technologies in the writing process}
During the preparation of this work, the author(s) used the ChatGPT AI Tool in order to improve language and readability (language editing). After using this tool/service, the author(s) reviewed and edited the content as needed and take(s) full responsibility for the content of the publication.

\bibliographystyle{jasr-model5-names}
\bibliography{refs.bib}


\end{document}